\documentclass[twocolumn,aps,pra,showpacs]{revtex4}
\usepackage[T1]{fontenc}
\usepackage[latin1]{inputenc}
\usepackage{amsmath}
\usepackage{amssymb}
\usepackage{graphicx}
\usepackage{esint}

\makeatletter
\@ifundefined{textcolor}{}
{%
 \definecolor{BLACK}{gray}{0}
 \definecolor{WHITE}{gray}{1}
 \definecolor{RED}{rgb}{1,0,0}
 \definecolor{GREEN}{rgb}{0,1,0}
 \definecolor{BLUE}{rgb}{0,0,1}
 \definecolor{CYAN}{cmyk}{1,0,0,0}
 \definecolor{MAGENTA}{cmyk}{0,1,0,0}
 \definecolor{YELLOW}{cmyk}{0,0,1,0}
}

\makeatother

\makeatother

\begin{document}

\title{Probing the critical exponent of superfluid fraction in a strongly
interacting Fermi gas}

\author{Hui Hu$^{1}$ and Xia-Ji Liu$^{1}$}

\email{xiajiliu@swin.edu.au}

\affiliation{$^{1}$Centre for Atom Optics and Ultrafast Spectroscopy, Swinburne
University of Technology, Melbourne 3122, Australia}

\date{\today}
\begin{abstract}
We theoretically investigate the critical behavior of second sound
mode in a harmonically trapped ultracold atomic Fermi gas with resonant
interactions. Near the superfluid phase transition with critical temperature
$T_{c}$, the frequency or the sound velocity of second sound mode
depends crucially on the critical exponent $\beta$ of superfluid
fraction. In an isotropic harmonic trap, we predict that the mode
frequency diverges like $(1-T/T_{c})^{\beta-1/2}$ when $\beta<1/2$.
In a highly elongated trap, the speed of second sound reduces by a
factor $1/\sqrt{2\beta+1}$ from that in a homogeneous three-dimensional
superfluid. Our prediction could be readily tested by measurements
of second sound wave propagation in a setup such as that exploited
by Sidorenkov \textit{et al.} {[}Nature \textbf{498}, 78 (2013){]}
for resonantly interacting lithium-6 atoms, once the experimental
precision is improved.
\end{abstract}


\maketitle

\section{Introduction}

Superfluidity, a remarkable state of matter in which particles flows
with zero resistance, is a ubiquitous quantum phenomenon occurring
in diverse systems ranging from liquid helium, high-temperature superconductors,
to neutron stars \cite{IntroductionToSuperfluidity,TheoryOfQuantumLiquids}.
While at zero temperature all the particles in the system participate
into the superfluid motion, at finite temperatures because of thermal
excitations only a portion of particles - named as superfluid fraction
- behaves in such a way. The remaining particles comprise a normal
fluid component that behaves like an ordinary fluid \cite{Tisza1938,Landau1941}.
To characterize superfluidity, it is therefore crucial to understand
the superfluid fraction, which, unfortunately is notoriously difficult
to calculate microscopically for strongly interacting quantum systems,
especially near the superfluid phase transition. In this respect,
the recently realized ultracold atomic Fermi gases with controllable
interatomic interactions and external harmonic trapping potentials
\cite{Ohara2002,Giorgini2008}, known as a new type of strongly interacting
superfluid, provide unique opportunities to explore superfluidity
and understand superfluid fraction in the strongly interacting regime.
In this paper, we propose that the critical behavior of superfluid
fraction of a resonantly interacting atomic Fermi gas at unitarity
(where atoms occupying unlike spin states interact with an infinitely
large scattering length) could be well characterized through the measurement
of second sound propagation.

Second sound, as well as first sound, is a coupled oscillation of
the superfluid and normal fluid components at finite temperatures
\cite{Tisza1938,Landau1941}. In contrast to first sound, which is
an in-phase oscillation of the two components (i.e., density oscillation),
second sound is an out-of-phase oscillation (i.e., temperature or
entropy wave) and depends very sensitively on the superfluid fraction.
Therefore, it presents arguably the most dramatic manifestation of
superfluidity. Indeed, in superfluid helium the accurate determination
of superfluid fraction slightly below the lambda point is provided
by the measurement of second sound \cite{Dash1957}. Very recently,
for a resonantly interacting Fermi gas of lithium-6 atoms confined
in highly elongated harmonic traps, the superfluid fraction is qualitatively
extracted from the measurement of second sound velocity along the
weakly confined axial direction, as reported by Sidorenkov \textit{et
al.} \cite{Sidorenkov2013}. This milestone experiment already imposes
a grand challenge, since the theoretical predictions for the temperature
dependence of the superfluid fraction in the unitary limit are rather
incomplete \cite{Taylor2008,Watanabe2010,Salasnich2010,Baym2013}.
Our proposal, together with future second sound measurement with better
precision in such ultracold atomic systems, allows an accurate determination
of the critical behavior of the superfluid fraction just below the
superfluid phase transition.

Our main results are briefly summarized as follows. We consider both
isotropic and highly elongated harmonic traps. The latter situation
is exactly the setup exploited in the current experiment \cite{Sidorenkov2013}.
For isotropic traps, we find that slightly below the superfluid transition
temperature $T_{c}$ the mode frequency of the second sound diverges
as $(1-T/T_{c})^{\beta-1/2}$ if the critical exponent of the superfluid
fraction $\beta<1/2$. While for highly elongated traps, the speed
of second sound along the weakly confined direction reduces by a factor
$1/\sqrt{2\beta+1}$ from that in a three-dimensional free space.
In both cases, the sensitive dependence of the second sound mode on
the critical exponent leads to an accurate calibration of $\beta$.

\section{Two-fluid hydrodynamics}

First and second sound are well described by the equations of two-fluid
hydrodynamics first derived by Landau \cite{Landau1941}. As discussed
in the previous works \cite{Taylor2008,Taylor2005,Taylor2009}, in
the dissipationless regime the solutions of these hydrodynamic equations
with frequency $\omega$ at temperature $T$ can be derived by minimizing
a variational action, which, in terms of displacement fields $\mathbf{u}_{s}(\mathbf{r})$
and $\mathbf{u}_{n}(\mathbf{r})$, is given by,$ $
\begin{eqnarray}
\mathcal{S} & = & \frac{1}{2}\int d\mathbf{r}\left[\omega^{2}\left(\rho_{s0}\mathbf{u}_{s}^{2}+\rho_{n0}\mathbf{u}_{n}^{2}\right)-\frac{1}{\rho_{0}}\left(\frac{\partial P}{\partial\rho}\right)_{\bar{s}}\left(\delta\rho\right)^{2}\right.\nonumber \\
 &  & \left.-2\rho_{0}\left(\frac{\partial T}{\partial\rho}\right)_{\bar{s}}\delta\rho\delta\bar{s}-\rho_{0}\left(\frac{\partial T}{\partial\bar{s}}\right)_{\rho}\left(\delta\bar{s}\right)^{2}\right].\label{eq:action}
\end{eqnarray}
Here, $\rho_{s}(\mathbf{r})$ and $\rho_{n}(\mathbf{r})$ are the
superfluid and normal fluid densities for a gas with total mass density
$\rho(\mathbf{r})\equiv mn=\rho_{s}+\rho_{n}$. $P(\mathbf{r})$ is
the local pressure of the gas and $\bar{s}(\mathbf{r})=s/\rho$ is
the entropy per unit mass. $\delta\rho(\mathbf{r})=-\mathbf{\nabla}\cdot[\rho_{s0}\mathbf{u}_{s}+\rho_{n0}\mathbf{u}_{n}]$
and $\delta\bar{s}(\mathbf{r})=-\mathbf{u}_{n}\cdot\mathbf{\mathbf{\nabla}}\bar{s}_{0}+(\bar{s}_{0}/\rho_{0})\mathbf{\mathbf{\nabla}}\cdot[\rho_{s0}(\mathbf{u}_{s}-\mathbf{u}_{n})]$
are the density and entropy fluctuations, respectively. The displacement
fields are related to the superfluid and normal velocity fields by
$d\mathbf{u}_{s}/dt=\mathbf{v}_{s}$ and $d\mathbf{u}_{n}/dt=\mathbf{v}_{n}$.
The effect of the external harmonic trapping potential $V_{T}(\mathbf{r})=m\omega_{T}^{2}r_{\perp}^{2}/2+m\omega_{z}^{2}z^{2}/2$
enters Eq. (\ref{eq:action}) through the position dependent equilibrium
thermodynamic functions, which we have indicated by the subscript
``$0$''. For a resonantly interacting Fermi gas, all these thermodynamic
functions - except the superfluid density - are known to certain precision,
owing to the recent experimental analysis of the homogeneous equation
of state performed by the MIT team \cite{Ku2012} by using the universality
relations satisfied by the unitary Fermi gas \cite{Ho2004,Hu2007}.
Throughout the work, we calculate the trapped density profile and
thermodynamic functions based on the smoothed experimental MIT data
\cite{Ku2012} and the local density approximation (LDA) which amount
to setting a local chemical potential $\mu(\mathbf{r})=\mu-V_{T}(\mathbf{r})$,
where $\mu$ is the chemical potential at the trap center.

In the absence of the coupling term between density and entropy fluctuations
(i.e., $\rho_{0}(\partial T/\partial\rho)_{\bar{s}}=0$), Eq. (\ref{eq:action})
admits two decoupled solutions: a pure in-phase mode with the \textit{Ansatz}
$\mathbf{u}_{s}(\mathbf{r})=\mathbf{u}_{n}(\mathbf{r})=\mathbf{u}^{(1)}(\mathbf{r})$
and a pure out-of-phase mode with $\rho_{s0}\mathbf{u}_{s}^{(2)}(\mathbf{r})+\rho_{n0}\mathbf{u}_{n}^{(2)}(\mathbf{r})=0$,
which may be referred to as first and second sound, respectively.
These first and second sound modes are the exact variational solutions
for pure density {[}$\delta T(\mathbf{r})=0${]} and pure temperature
{[}$\delta\rho(\mathbf{r})=0${]} oscillations. When the coefficient
$\rho_{0}(\partial T/\partial\rho)_{\bar{s}}$ is nonzero, the first
and second sound are necessarily coupled. This coupling can be conveniently
characterized by the dimensionless Landau-Placzek (LP) parameter $\epsilon_{\textrm{LP}}\equiv\gamma-1$
\cite{Hu2010}, where $\gamma\equiv\bar{c}_{p}/\bar{c}_{v}$ is the
ratio between the equilibrium specific heats per unit mass at constant
pressure {[}$\bar{c}_{p}=T(\partial\bar{s}/\partial T)_{P}${]} and
density {[}$\bar{c}_{v}=T(\partial\bar{s}/\partial T)_{\rho}${]}.
In superfluid helium, $\bar{c}_{p}\simeq\bar{c}_{v}$ or $\epsilon_{\textrm{LP}}\simeq0$,
implying $\rho_{0}(\partial T/\partial\rho)_{\bar{s}}\simeq0$. Thus,
the solutions of the two-fluid hydrodynamic equations for superfluid
helium are perfectly described by decoupled first and second sound
modes. For resonantly interacting atomic Fermi gases, the universality
relations give rise to $\rho_{0}(\partial T/\partial\rho)_{\bar{s}}=2T/3\neq0$
\cite{Taylor2008}. Close to the superfluid transition temperature
$T_{c}\simeq0.167T_{F}$, where $T_{F}$ is the Fermi temperature,
we estimate from the MIT data that the LP parameter is about $\epsilon_{\textrm{LP}}\simeq0.4$.
Therefore, similarly to superfluid liquid helium, the solutions of
two-fluid equations for a unitary Fermi gas are well approximated
by weakly coupled first and second sound modes. We note that the smallness
of the LP parameter in superfluid helium and unitary Fermi gas and
hence the weak coupling between their first and second modes is a
general consequence of strong interactions \cite{Hu2010}. Note also
that the existence of harmonic traps will significantly reduce the
sound mode coupling, as we shall see later. Hereafter, we focus on
the second sound mode by neglecting its coupling to the first sound
mode. In the past, the first sound mode of a unitary Fermi gas has
been studied in greater detail, both at zero temperature \cite{Stringari2004,Hu2004,Altmeyer2007,Joseph2007}
and finite temperatures \cite{Tey2013}.

\section{Second sound near superfluid transition}

Inserting the \textit{Ansatz} $\mathbf{u}_{n}^{(2)}(\mathbf{r})=-(\rho_{s0}/\rho_{n0})\mathbf{u}_{s}^{(2)}(\mathbf{r})$
for second sound into Eq. (\ref{eq:action}), taking the variation
with respect to $\mathbf{u}_{s}^{(2)}(\mathbf{r})$ and making use
of standard thermodynamic relations, we obtain the following equation
for the superfluid displacement field \cite{Taylor2009}:
\begin{equation}
\omega^{2}\mathbf{u}_{s}^{(2)}=-\bar{s}_{0}\mathbf{\nabla}\left[\frac{1}{\rho_{0}}\left(\frac{\partial T}{\partial\bar{s}}\right)_{\rho}\mathbf{\nabla}\cdot\left(\frac{s_{0}\rho_{s0}}{\rho_{n0}}\mathbf{u}_{s}^{(2)}\right)\right].\label{eq:secondsound1}
\end{equation}
As $\delta\rho(\mathbf{r})=0$ for second sound, we may also rewrite
the above equation into a closed form for the temperature fluctuation
$\delta T(\mathbf{r})=(\partial T/\partial\rho)_{\bar{s}}\delta\rho(\mathbf{r})+(\partial T/\partial\bar{s})_{\rho}\delta\bar{s}(\mathbf{r})=\rho_{0}^{-1}(\partial T/\partial\bar{s})_{\rho}\mathbf{\nabla}\cdot[s_{0}\rho_{s0}\mathbf{u}_{s}^{(2)}/\rho_{n0}]$,
where $\delta\bar{s}(\mathbf{r})$ is the entropy fluctuation described
below Eq. (\ref{eq:action}). This gives rise to,
\begin{equation}
\omega^{2}\delta T\left(\mathbf{r}\right)=-\frac{1}{\rho_{0}}\left(\frac{\partial T}{\partial\bar{s}}\right)_{\rho}\mathbf{\nabla\cdot}\left[\frac{\bar{s}_{0}^{2}\rho_{0}\rho_{s0}}{\rho_{n0}}\mathbf{\nabla}\delta T\left(\mathbf{r}\right)\right].\label{eq:secondsound2}
\end{equation}
In homogeneous space, where $\delta T\left(\mathbf{r}\right)\propto e^{i\mathbf{k}\cdot\mathbf{r}}$,
we recover the well-known result for the second sound velocity: $c_{2,hom}^{2}=T(\bar{s}_{0}^{2}\rho_{s0})/(\bar{c}_{v}\rho_{n0})$.

In the presence of harmonic traps, it seems to be cumbersome to directly
solve Eq. (\ref{eq:secondsound2}). Fortunately, near the superfluid
transition this equation could be greatly simplified, as the temperature
oscillation has to be restricted in a small superfluid area around
the trap center and therefore we may safely neglect the position dependence
of all thermodynamic functions - except the superfluid density. Furthermore,
close to transition it is reasonable to assume the following critical
behavior for superfluid fraction,
\begin{equation}
\frac{\rho_{s0}}{\rho_{n0}}(T)\simeq\eta\left(1-\frac{T}{T_{c}}\right)^{2\beta},\label{eq:superfluidfraction}
\end{equation}
where the constant $\eta$ and the critical exponent $\beta$ are
to be determined for a unitary Fermi gas. It is known from the Leggett
model of pairing at unitarity that the mean-field BCS wave function
gives rise to $\eta=2$ and $\beta=1/2$ \cite{QuantumLiquids}. However,
a superfluid with a two-component order parameter (and a bosonic fluctuation
spectrum) generally undergoes a second order phase transition with
a superfluid density that varies as $\rho_{s}\propto(T_{c}-T)^{2/3}$
close to the transition, independent of the interaction strength \cite{StatisticalPhysics}.
Indeed, in superfluid liquid helium, the second sound measurement
suggests that $\eta\simeq3.2$ and $\beta\simeq1/3$ \cite{Dash1957}.
With these considerations, we find that within LDA:
\begin{eqnarray}
\omega^{2}\delta T & = & -\frac{\eta k_{B}T_{F}f_{s}^{2}}{mf_{s}^{'}}\mathbf{\nabla\cdot}\left\{ \left(1-\frac{T}{T_{c}}\right)^{2\beta}\right.\times\nonumber \\
 &  & \left.\left[1-\frac{TV_{T}\left(\mathbf{r}\right)/k_{B}}{\left(T_{c}-T\right)\left(T_{F}f_{\mu}-Tf_{\mu}^{'}\right)}\right]^{2\beta}\mathbf{\nabla}\delta T\right\} ,\label{eq:secondsound3}
\end{eqnarray}
where we have expressed the entropy and chemical potential of a uniform
unitary Fermi gas in terms of dimensionless functions of the reduced
temperature $T/T_{F}$: $s=nk_{B}f_{s}(T/T_{F})$ and $\mu=k_{B}T_{F}f_{\mu}(T/T_{F})$;
$T_{F}=\hbar^{2}(3\pi^{2}n)^{2/3}/(2mk_{B})$ is the Fermi temperature
at the trap center with density $n$, and $f_{s,\mu}^{'}\equiv df_{s,\mu}/d(T/T_{F})$.

\begin{figure}
\begin{centering}
\includegraphics[clip,width=0.48\textwidth]{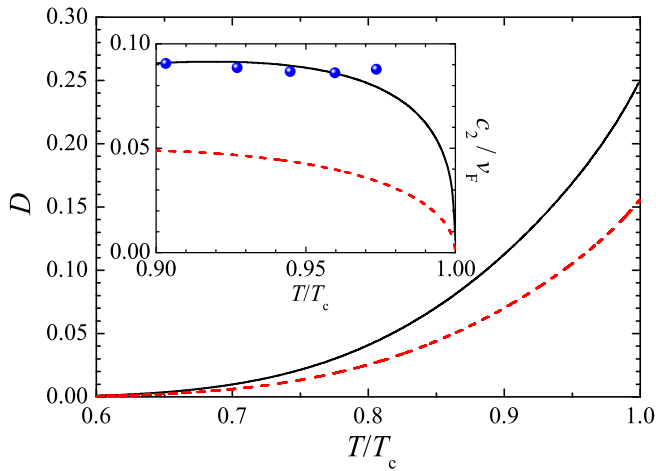} 
\par\end{centering}

\caption{(Color online) Temperature dependence of the parameter $\mathcal{D}$
near the superfluid transition temperature $T_{c}\simeq0.167T_{F}$,
for a strongly interacting unitary Fermi gas. We consider two types
of critical behavior for the superfluid fraction $\eta(1-T/T_{c})^{2\beta}$:
(1) superfluid helium for which $\eta\simeq3.2$ and $\beta\simeq1/3$,
as shown by the black solid line, and (2) mean-field theory in which
$\eta=2$ and $\beta=1/2$, as shown by the red dashed line. The inset
shows the second-sound velocity of a homogeneous unitary Fermi gas
$c_{2,hom}=\sqrt{T(\bar{s}_{0}^{2}\rho_{s0})/(\bar{c}_{v}\rho_{n0})}$
in units of the Fermi velocity $v_{F}=\hbar k_{F}/m$, calculated
using the assumed superfluid fraction Eq. (\ref{eq:superfluidfraction}).
The blue solid circles are the theoretical predictions obtained by
using the measured superfluid density, which was found to be close
to that of superfluid helium \cite{Sidorenkov2013}.}

\label{fig1} 
\end{figure}

It is readily seen that the superfluid area is restricted to $r_{\perp}\leq R_{\perp s}$
and $z\leq R_{zs}$, where $m\omega_{T}^{2}R_{\perp s}^{2}/2=m\omega_{z}^{2}R_{zs}^{2}/2=(T_{c}/T-1)k_{B}(T_{F}f_{\mu}-Tf_{\mu}^{'})$.
By introducing the scaled coordinates $\tilde{r}_{\perp}=r_{\perp}/R_{\perp s}$
and $\tilde{z}=z/R_{\perp s}$, we may rewrite Eq. (\ref{eq:secondsound3})
into the following dimensionless form,
\begin{equation}
\tilde{\omega}^{2}\delta T+\tilde{\mathbf{\nabla}}\cdot\left[\left(1-\tilde{r}_{\perp}^{2}-\tilde{z}^{2}/\lambda^{2}\right)^{2\beta}\tilde{\mathbf{\nabla}}\delta T\right]=0,\label{eq:secondsound4}
\end{equation}
where $\lambda\equiv\omega_{T}/\omega_{z}$ is the aspect ratio of
the harmonic trap and a reduced mode frequency $\tilde{\omega}$ is
defined by 
\begin{equation}
\omega\equiv\tilde{\omega}\sqrt{\mathcal{D}}(1-T/T_{c})^{\beta-1/2}\omega_{T}\label{eq:reducedfrequency}
\end{equation}
with $\mathcal{D}=\eta(TT_{F}f_{s}^{2})/[2T_{c}f_{s}^{'}(T_{F}f_{\mu}-Tf_{\mu}^{'})]$.
In Fig. 1, we show the temperature dependence of the parameter $\mathcal{D}$
near the superfluid transition, calculated using the MIT data for
the equation of state of a unitary Fermi gas \cite{Ku2012}. It is
typically at about $0.2$. From Eq. (\ref{eq:reducedfrequency}),
one may realize immediately that for any non-zero discrete mode frequency,
it would become divergent when temperature approaches to the superfluid
transition temperature if the critical exponent $\beta<1/2$. 

\begin{figure}
\begin{centering}
\includegraphics[clip,width=0.48\textwidth]{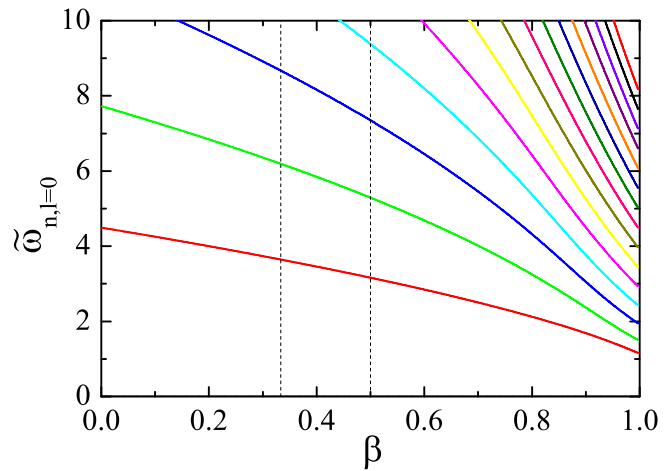} 
\par\end{centering}

\caption{(Color online) Reduced second-sound mode frequency as a function of
the critical exponent $\beta$, for an isotropically trapped unitary
Fermi gas. We consider the $l=0$ sector only. In the cases that $\beta$
is an integer or half-integer, the reduced frequency is known analytically.
In particular, the reduced mode frequency at $\beta=1/2$ is given
by $\tilde{\omega}_{n,l=0}=2\sqrt{n(n+3/2)}.$}

\label{fig2} 
\end{figure}

\subsection{Isotropic traps}

For an isotropic harmonic trap, we may recast Eq. (\ref{eq:secondsound4})
into a one-dimensional differential equation, for example, in the
sector of zero angular momentum $l=0$ (i.e., breathing modes), 
\begin{equation}
\left[R\left(x\right)\frac{d^{2}}{dx^{2}}+P\left(x\right)\frac{d}{dx}+Q\left(x\right)\right]\delta T\left(x\right)=0,\label{eq:isotropic1D}
\end{equation}
where $x\equiv\tilde{r}^{2}\leq1$ and the coefficients $R(x)=x(1-x)^{2\beta+1}$,
$P(x)=[3/2-(3/2+2\beta)x](1-x)^{2\beta}$ and $Q(x)=\tilde{\omega}^{2}(1-x)/4$.
It can be solved numerically by using a multi-series expansion method
\cite{Liu2008}. The numerical results are reported in Fig. 2. The
reduced mode frequencies decrease quickly with increasing the critical
exponent $\beta$. When $\beta$ is an integer or half-integer, our
numerical results could be examined analytically, as the solutions
for the temperature fluctuation are simply polynomials and therefore
the mode frequency are known precisely. Indeed, for $\beta=1/2$ (i.e.,
mean-field superfluid fraction), Eq. (6) has exactly the same structure
as the hydrodynamic equation that describes collective oscillations
of a zero-temperature Bose condensate \cite{Stringari1996,Fliesser1997}.
It admits analytical solutions for harmonic traps with arbitrary aspect
ratio. In the case of isotropic traps, the reduced mode frequency
is given by $\tilde{\omega}_{nl}=2\sqrt{n^{2}+nl+3n/2+l/2}$ \cite{Stringari1996}.

\begin{figure}
\begin{centering}
\includegraphics[clip,width=0.48\textwidth]{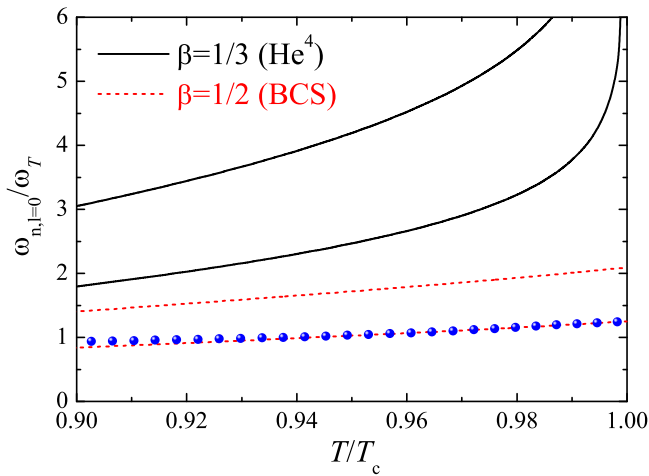} 
\par\end{centering}

\caption{(Color online) Temperature dependence of the two (lowest breathing)
second-sound mode frequency of an isotropically trapped unitary Fermi
gas near the superfluid transition temperature. For a critical exponent
$\beta<1/2$, the mode frequency diverges at the transition. The blue
solid circles are the full variational results of the two-fluid hydrodynamic
equations for the lowest breathing second-sound mode. }

\label{fig3} 
\end{figure}

Now, we are able to calculate the mode frequency, by using Eq. (\ref{eq:reducedfrequency}).
The two lowest mode frequencies are shown in Fig. \ref{fig3} for
superfluid-helium-like superfluid fraction (black solid lines) or
mean-field-like superfluid fraction (red dashed lines). In the former
case, the divergence of the mode frequency near superfluid transition
is evident, as we may anticipate. 

For isotropic traps, it is worth noting that the equations of two-fluid
hydrodynamics can be fully solved by using a variational approach
\cite{Taylor2009}. We have preformed such a calculation with a mean-field
superfluid fraction. The results for the lowest second-sound mode
frequency are reported in Fig. \ref{fig3} by blue solid circles.
The good agreement between the full variational calculation and the
prediction of the simplified second-sound Eq. (\ref{eq:secondsound4})
gives a reasonable justification for all the assumptions that we have
made to derive Eq. (\ref{eq:secondsound4}), including neglecting
the coupling between first and second sound modes and the ignorance
of the position dependence of thermodynamic functions used in Eq.
(\ref{eq:secondsound2}).

\begin{figure}
\begin{centering}
\includegraphics[clip,width=0.48\textwidth]{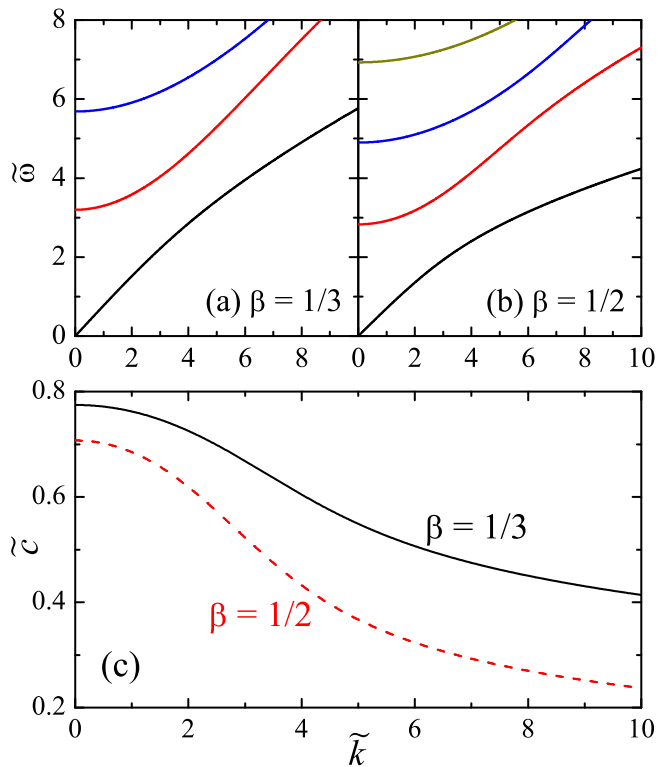} 
\par\end{centering}

\caption{(Color online) Reduced second-sound mode frequency for a unitary Fermi
gas confined in highly elongated harmonic traps for the critical exponent
$\beta=1/3$ (a) and $\beta=1/2$ (b). We assume that the second sound
can propagate freely along the long trap axis. In (c), we show the
reduced second-sound velocity $\tilde{c}=\partial\tilde{\omega}/\partial\tilde{k}$
for the lowest phonon modes.}

\label{fig4} 
\end{figure}

\subsection{Highly elongated traps}

Experimentally, a resonantly interacting atomic Fermi gas is trapped
in highly elongated harmonic trapping potentials. The second-sound
is excited and propagated one-dimensionally along the long trap direction
\cite{Sidorenkov2013}. For such a configuration, we may assume that
the temperature fluctuation has the form: $\delta T(\tilde{r}_{\perp},\tilde{z})=\delta T(\tilde{r}_{\perp})e^{i\tilde{k}\tilde{z}},$
where $\tilde{k}\equiv kR_{\perp s}$ is the reduced wave vector.
The fluctuation field $\delta T(\tilde{r}_{\perp})$ then satisfies,
\begin{equation}
\left[\tilde{\omega}^{2}-\tilde{k}^{2}\left(1-\tilde{r}_{\perp}^{2}\right)^{2\beta}\right]\delta T+\tilde{\mathbf{\nabla}}\cdot\left[\left(1-\tilde{r}_{\perp}^{2}\right)^{2\beta}\tilde{\mathbf{\nabla}}\delta T\right]=0.\label{eq:secondsound5}
\end{equation}
For any reduced wave vector $\tilde{k}$, similar to the case of isotropic
traps, the above equation in the sector of $l=0$ can be rewritten
in a one-dimensional (1D) differential form Eq. (\ref{eq:isotropic1D})
by setting $x=\tilde{r}_{\perp}^{2}$ , but with new coefficients
$P(x)=[1-(1+2\beta)x](1-x)^{2\beta}$ and $Q(x)=[\tilde{\omega}^{2}-(1-x)^{2\beta}](1-x)/4$.
It can be solved numerically following Ref. \cite{Liu2008}. Fig.
\ref{fig4} shows the results for $\beta=1/3$ and $\beta=1/2$. In
the latter case, our result in Fig. \ref{fig4}(b) agrees exactly
with earlier prediction on first-sound propagation of a zero-temperature
Bose condensate in highly elongated harmonic traps \cite{Zaremba1998,Cappuzzi2006,Ghosh2006},
as it should be. 

In Fig. \ref{fig4}, we observe multi-branches in the spectrum, each
of which corresponds to a discrete radial excitation \cite{Zaremba1998,Ghosh2006}.
The lowest branch is of particular interest, as it resembles the phonon
mode in free space and is the easiest mode to excite experimentally.
The associated second-sound velocity is given by, $ $$c_{2,1D}=\partial\omega/\partial k=(\partial\tilde{\omega}/\partial\tilde{k})c_{2,hom}$.
Thus, the velocity of second-sound propagated in quasi-1D geometry
is reduced by a factor of $\tilde{c}=\partial\tilde{\omega}/\partial\tilde{k}$,
with respect to the bulk value $c_{2,hom}$. The similar quenching
of first-sound velocity due to confinement was pointed out earlier
for a unitary Fermi gas \cite{Cappuzzi2006} or a Bose condensate
\cite{Zaremba1998}. The value of $\tilde{c}$ may be calculated by
integrating out the transverse coordinate $\tilde{r}_{\perp}$ in
Eq. (\ref{eq:secondsound5}): $\tilde{c}^{2}=\int d\mathbf{r}_{\perp}(1-\tilde{r}_{\perp}^{2})^{2\beta}\delta T(\tilde{r}_{\perp})/\int d\mathbf{r}_{\perp}\delta T(\tilde{r}_{\perp})$.
In the limit of long wave length ($\tilde{k}\rightarrow0$), where
the temperature fluctuation $\delta T(\tilde{r}_{\perp})$ is radially
independent, we find that $\tilde{c}=1/\sqrt{2\beta+1}$, in agreement
with our numerical result shown in Fig. \ref{fig4}(c).

\begin{figure}
\begin{centering}
\includegraphics[clip,width=0.48\textwidth]{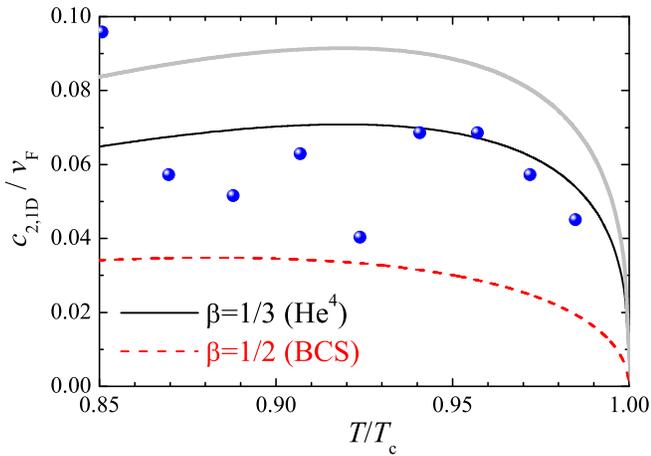} 
\par\end{centering}

\caption{(Color online) Second-sound velocity for a unitary Fermi gas confined
in highly elongated harmonic traps for the critical exponent $\beta=1/3$
(black solid line) and $\beta=1/2$ (red dashed line). The blue solid
circles are the experimental results extracted from Fig. 3(a) of Ref.
\cite{Sidorenkov2013}, by assuming that the peak density at the trap
center is unchanged close to superfluid transition. For comparison,
we also show by the gray line the bulk second-sound velocity for $\beta=1/3$.}

\label{fig5} 
\end{figure}

In Fig. \ref{fig5}, we report the 1D second-sound velocity $c_{2,hom}/\sqrt{2\beta+1}$
by using the assumed superfluid-helium or mean-field like superfluid
fraction Eq. (\ref{eq:superfluidfraction}). For comparison, we show
also the experimental data extracted from Fig. 3(a) of Ref. \cite{Sidorenkov2013}.
Close to the superfluid transition (i.e., $T>0.95T_{c}$), our prediction
of the 1D second-sound velocity with superfluid-helium-like superfluid
fraction agrees reasonably well with the measurement. By noting that
the superfluid fraction of a unitary Fermi gas resembles that of superfluid
helium \cite{Sidorenkov2013}, this agreement somehow is an indication
of the quenched second sound velocity. However, a quantitative experimental
determination of the critical exponent $\beta$ requires a much better
precision of data.

\subsection{Conclusions}

In conclusion, we have investigated theoretically how the second sound
of a harmonically trapped unitary Fermi gas is affected by the critical
exponent $\beta$ of superfluid fraction near the superfluid phase
transition when temperature $T$ approaches to the critical temperature
$T_{c}$. In an isotropic trap , the sound frequency goes like $(1-T/T_{c})^{\beta-1/2}$
and therefore exhibit clearly a divergence when $\beta<1/2$. In an
experimentally exploited highly elongated trap, the second sound velocity
along the long trap axis reduces by a factor $1/\sqrt{2\beta+1}$
with respect to its bulk value. Our prediction could be used to measure
directly the critical exponent $\beta$ in future experiments, if
the experimental accuracy gets improved.
\begin{acknowledgments}
The present work is dedicated to in memory of Professor Allan Griffin,
who was always enthusiastic on observing the second sound in resonantly
interacting atomic Fermi gases and was the driving force of our recent
works \cite{Taylor2008,Taylor2009,Hu2010}. We thank Edward Taylor,
Allan Griffin, Lev Pitaevskii and Sandro Stringari for their stimulating
discussions during the early stage of this work in 2009. This work
is supported by the ARC Discovery Projects (Grant No. DP0984522 and
DP0984637) and NFRP-China (Grant No. 2011CB921502).\end{acknowledgments}

\end{document}